\documentclass[fleqn,twoside]{article}
\usepackage{espcrc2}

\usepackage{graphicx}
\usepackage[figuresright]{rotating}

\newcommand{\AmS}{{\protect\the\textfont2
  A\kern-.1667em\lower.5ex\hbox{M}\kern-.125emS}}

\title{Leptogenesis and Low-energy Observables}

\author{Gustavo C. Branco \address[MCSD]{Departamento de 
F\'{\i}sica and Centro de F\'{\i}sica Te\'orica de
Part\'{\i}culas (CFTP),
Instituto Superior T\'ecnico, Av. Rovisco Pais, P-1049-001 Lisboa, Portugal}
        \thanks{Invited talk presented by G.C.B. Email: gbranco@ist.utl.pt},
         and
        M. N. Rebelo\addressmark[MCSD]\thanks{Email: rebelo@ist.utl.pt}}
       
\begin{document}

\begin{abstract}
We address the question of how to establish a connection between 
leptogenesis and low energy observables. We emphasize that such a 
connection only exists in the framework of flavour models. 
A particular example is the case of texture zeros in some of the 
Yukawa couplings.
\vspace{1pc}
\end{abstract}

\maketitle

\section{Introduction}
The experimental discovery of neutrino masses and leptonic
mixing is one of the most important recent developments in
Particle Physics. It is well known that in the 
Standard Model (SM), neutrinos are strictly massless,
due to the absence of right-handed neutrinos, together
with exact $B-L$ conservation. Therefore, the discovery of 
non-vanishing neutrino masses is a clear indication of
physics beyond the SM.

One of the simplest ways of accommodating neutrino masses
is through the introduction of at least two right-handed 
neutrinos, leading to the seesaw mechanism. Apart from 
offering an elegant explanation for the smallness of 
neutrino masses, the seesaw mechanism also provides a 
framework to create the observed baryon asymmetry of the 
Universe (BAU). This is achieved through leptogenesis, a
mechanism first suggested in Ref.~\cite{Fukugita:1986hr}, 
where out-of-equilibrium
decays of heavy right-handed neutrinos create a lepton 
asymmetry which in turn is converted into a baryon asymmetry,
through sphaleron interactions \cite{Klinkhamer:1984di},
\cite{Kuzmin:1985mm}. 
Since leptogenesis, together with
neutrino masses, leptonic mixing and 
leptonic CP violation, all arise from the seesaw mechanism,
it is natural to ask whether there is a connection between leptogenesis
and low energy observables. Furthermore, such a connection 
would be necessary in order to test leptogenesis, assuming the
heavy neutrinos masses lie outside the reach of future
experiments.

\section{Low-energy observables}

Let us assume that lepton number is violated at a high energy scale,
leading to the generation at low energies of an effective
left-handed Majorana neutrino mass matrix. In the mass eigenstate 
basis, the charged weak current can be written:
\begin{equation}
{\cal L}_W = - \frac{g}{\sqrt{2}}  \overline{l_{jL}}
\gamma_{\mu} U_{jk} {\nu_k}_L + h.c.,
\label{lw}
\end{equation}
where $U$ denotes the leptonic mixing matrix at low energies,
usually named the Pontecorvo, Maki, Nakagawa and Sakata (PMNS)
matrix. Although in the seesaw framework $U$ is not exactly unitary, 
in the standard seesaw type I framework, $U$ is unitary to a high 
degree of accuracy. Once $3 \times 3$ unitarity is assumed, $U$ 
is caracterized by six parameters which are usually taken as three 
mixing angles and three CP violating phases. Due to the assumed
Majorana nature of neutrinos, the simplest rephasing 
invariant functions of $U_{ij}$ are the bilinears
$U_{l\alpha } U^*_{l\beta }$ (no summation on repeated indices implied).
Using unitarity, it has been shown \cite{Branco:2008ai}
that the full matrix $U$ can be 
constructed from six independent Majorana-type phases
$\phi_{\alpha \beta} \equiv \arg \ (U_{l\alpha } U^*_{l\beta })$.
There are nine low energy observables, namely the three light
neutrino masses and the six parameters characterizing $U$
\cite{Mohapatra:2005wg}.
The question is then whether it is possible to relate leptogenesis
to the low energy observables.

\section{Leptogenesis and the relation to Low-energy Observables}
Let us consider the SM with the addition of three 
right-handed (r.h.) neutrinos.
In this case, one can write an $SU(2) \times U(1)$ invariant
Majorana mass term for r.h. neutrinos, denoted $M_R$, assumed 
to be of a scale much higher than the electroweak scale, $v$.
After spontaneous  $SU(2) \times U(1)$ breaking, a neutrino 
Dirac mass matrix $m_D$ is also generated. This leads 
to an effective $3 \times 3$ Majorana mass matrix 
$m_{eff}= - m_D \frac{1}{M_R}{m_D}^T$.
The full leptonic mixing matrix can then be written:
\begin{eqnarray}
{\cal L}_W = & - \frac{g}{\sqrt{2}} & \left( \overline{l_{iL}}
\gamma_{\mu} K_{ij} {\nu_j}_L +
\overline{l_{iL}} \gamma_{\mu} G_{ij} {N_j}_L \right) W^{\mu} 
\nonumber \\
& +h.c. &
\label{phys}
\end{eqnarray}
where $K$ and $G$ are the first three rows of the $6 \times 6$
unitary matrix which diagonalizes the full neutrino mass matrix
(which includes both $m_D$ and $M_R$), in the weak basis (WB) 
where the charged
lepton mass matrix is already real and diagonal. 
It can be shown that, to an
excellent, approximation $G = m_D D^{-1}$ in the WB
where $M_R$ is also diagonal and real and 
where $D =$ diag$(M_1, M_2, M_3)$ with $M_i$ denoting
the masses of the three heavy neutrinos $N_i$.  
$K$ coincides very approximately, up to corrections of order
$v/ M_R $, with the unitary matrix
that diagonalizes $m_{eff}$ in the WB where the charged lepton 
mass matrix is diagonal real, i.e., the PMNS matrix. 

The lepton number asymmetry generated through CP violating decays of
the j-th  heavy Majorana neutrino into the different leptonic
families has been computed \cite{ref-1,ref0,ref1,ref2,ref3} in the
single flavour approximation, and shown to be proportional
to:
\begin{equation}
A^j \, \alpha \, \sum_{k \ne j} 
{\rm Im} ({m_D}^\dagger m_D)_{jk} ({m_D}^\dagger m_D)_{jk}
\end{equation}
in the weak basis (WB) where $M_R$ is diagonal.
In the seesaw framework the matrix $m_D$, in this WB, can be written  
\cite{Casas:2001sr}:
\begin{equation}
m_D = i U {\sqrt d} R {\sqrt D}
\end{equation}
where $D$ stands for $M_R$ and $R$ is a general complex orthogonal matrix.
The matrix $R$ is relevant for leptogenesis since:
\begin{equation}
m^\dagger_D m_D = - \sqrt{D} R^\dagger d R
\sqrt{D}.  
\label{drr}
\end{equation}
Eq.~(\ref{drr}) shows that in general 
unflavoured leptogenesis is independent of
the presence of CP violation at low energies  \cite{Rebelo:2002wj}.
Notice that in the WB where
the charged lepton mass matrix and $M_R$ are real and diagonal,
all CP violating phases appear in $m_D$. 

It is possible to impose constraints on $m_D$ in the context
of special flavour models. A particular example 
is the imposition of zero textures on $m_D$ in the WB
where the charged lepton mass matrix and $M_R$ are 
real and diagonal. In this case one has
\begin{equation}
(m_{D})_{ij}=0~ \Rightarrow ~ (U)_{ik}\sqrt{d}_{kk}R_{kj}=0.  
\label{orto}
\end{equation}
corresponding to an orthogonality condition between one column of
the matrix $R$ and one row of the matrix $U \sqrt{d}$.
Imposing texture zeros on $m_D$ reduces the number of
CP violating phases and in some cases, allows to fully relate
the matrix $R$ to low energy parameters while at the same
time imposing constraints on 
low energy physics  \cite{Branco:2005jr,Branco:2007nb}.
The expression written above relating the mixing matrix $G$
to the matrix $m_D$ also shows how
constraints on $m_D$ may affect high energy physics. 

It is important to notice that textures are not WB independent
and symmetries are only explicit in specially chosen WB.
On the other hand it may be convenient to analyse specific
flavour models without the  requirement of being
in a particular WB or going to the physical basis.  
CP-odd WB invariants can be useful to study CP violation both in the
quark and leptonic sector. The technique to obtain 
CP-odd WB invariant conditions  was developed for the
first time in   \cite{Bernabeu:1986fc}  for the study of CP violation
in the quark sector of the Standard Model. This is a powerful tool,
and the fact that constraints imposed on the mass matrices
will in general reduce the number of CP violating phases, 
also allows to use this type of conditions in particular cases, to
recognize  models with texture zeros
when written in a WB where these are not present \cite{Branco:2005jr}.
Weak basis invariant conditions relevant for CP violation in the 
leptonic sector at low energies must be sensitive to the
Dirac-type phase as well as to Majorana-type phases. These were
given in \cite{Branco:1986gr,Branco:1998bw,Branco:2001pq}.
Weak basis invariant condititions relevant in the case of unflavoured 
leptogenesis were first presented in \cite{Pilaftsis:1997jf},
\cite{Branco:2001pq}. Details and further analysis are given in 
\cite{Branco:2004hu}, \cite{Dreiner:2007yz}.

Leptogenesis in the single flavour approximation
relies on the assumption that washout
effects are not sensitive to the different flavours of
charged leptons into which the heavy neutrino decays.
It was pointed out that flavour matters in leptogenesis
whenever the mass of the lightest heavy neutrino
is lower than $10^{12}$ GeV   
\cite{Barbieri:1999ma,Endoh:2003mz,Fujihara:2005pv,Abada:2006fw,Nardi:2006fx,Abada:2006ea}.

The separate lepton $i$ family asymmetry generated from the decay
of the $k$th heavy Majorana neutrino depends on the combination
\cite{Fujihara:2005pv}
Im$\left( (m_D^\dagger m_D)_{k k^\prime}(m_D^*)_{ik} (m_D)_{ik^\prime}\right) $
as well as on
Im$\left( (m_D^\dagger m_D)_{k^\prime k}(m_D^*)_{ik} (m_D)_{ik^\prime}\right) $
It is clear from these expressions that in the case of flavoured
leptogenesis there are additional sources of CP violation
since the PMNS matrix does not cancel out. This gives rise to
the new possibility of having viable leptogenesis even in the case
of $R$ being a real matrix. For some of the early attempts see  
\cite{Branco:2006hz,Pascoli:2006ie,Branco:2006ce,Uhlig:2006xf}
In the case of real $R$, the phases relevant for leptogenesis 
are arguments of the Majorana bilinears introduced in section 2. 

\section*{Acknowledgements}
This work was partially supported by Funda\c c\~ ao para a
Ci\^ encia e a  Tecnologia (FCT, Portugal) through the projects
POCI/81919/2007, CERN/FP/83503/2008
and CFTP-FCT UNIT 777  which are partially funded through POCTI
(FEDER) and by the Marie Curie RTN MRTN-CT-2006-035505. G.C.B. 
would like to thank G.L. Fogli, E. Lisi
and their colleagues in the organizing committee 
of Now 2008  for warm hospitality at this very stimulating Meeting.

\end{document}